\documentclass[12pt,twoside]{article}
\usepackage{amsmath, amsfonts, amscd, amssymb}
\usepackage{amsthm}
\usepackage{fancybox, eufrak, euscript, oldgerm}
\newcommand{\aconn}{\mathcal{A}}

\newcommand{\aut}{{\mathcal{A}}ut}

\newcommand{\conn}{\mathcal{D}}

\newcommand{\curv}{R}

\newcommand{\ricci}{\EuScript{R}}

\newcommand{\diff}{\mathbf{\Omega}}

\newcommand{\gauge}{\mathcal{U}}

\newcommand{\modl}{\mathbf{\mathcal{E}}}

\newcommand{\Omg}{\mathbf{\Omega}}

\newcommand{\smooth}{\mathcal{C}^{\infty}}
\newcommand{\sconn}{\textsf{A}}

\newcommand{\struc}{\mathbf{A}}

\newcommand{\mapto}{\longrightarrow}

\title{\underline{Glafka--2004: `2-in-1'}\\{\Large\bf I. `Iconoclasm': The Soul and Spirit of the Meeting}\\
\underline{\large\sl and}\\ {\Large\bf II. Categorical Quantum
Gravity}\thanks{In Part {\bf I} the introductory talk to the 1st
{\itshape Glafka--2004: Iconoclastic Approaches to Quantum
Gravity} international theoretical physics conference, held in
Athens (Greece, summer 2004) is given. In Part {\bf II} this
author's more technical talk at the conference is presented in
paper form. The original title of the technical talk at the
conference was ``{\em Abstract Differential Geometric Excursion to
Classical and Quantum Gravity}'', which has been shortened here.
Both papers, but separately, are destined to appear in a special
proceedings issue of the International Journal of Theoretical
Physics (Ioannis Raptis, Guest Editor).}}

\author{Ioannis Raptis (Glafka Scientific Organizer)\thanks{European Commission Marie Curie Reintegration Research Fellow,
Algebra and Geometry Section, Department of Mathematics,
University of Athens, Panepistimioupolis, Athens 157 84, Greece;
\underline{\em and} Visiting Researcher, Theoretical Physics
Group, Blackett Laboratory, Imperial College of Science,
Technology and Medicine, Prince Consort Road, South Kensington,
London SW7 2BZ, UK; e-mail: i.raptis@ic.ac.uk}}

\date{\today}

\begin{document}

{\catcode`\ =13\global\let =\ \catcode`\^^M=13
\gdef^^M{\par\noindent}}
\def\verbatim{\tt
\catcode`\^^M=13 \catcode`\ =13 \catcode`\\=12 \catcode`\{=12
\catcode`\}=12 \catcode`\_=12 \catcode`\^=12 \catcode`\&=12
\catcode`\~=12 \catcode`\#=12 \catcode`\%=12 \catcode`\$=12
\catcode`|=0 }

\maketitle

\pagestyle{myheadings}\markboth{\centerline {\small {\sc {Ioannis
Raptis}}}}{\centerline {\footnotesize {\sc {Glafka 2-in-1:
`Iconoclastic QG' and `Categorical QG'}}}}

\pagenumbering{arabic}

\begin{abstract}

\noindent The following is a 2-part, `2-in-1' paper. In part I, a
brief talk that opened and attempted to set the atmosphere for the
first `{\itshape Glafka--2004: Iconoclastic Approaches to Quantum
Gravity}' international theoretical physics conference is
presented in paper form. The talk aimed to capture the general
spirit of the meeting, as well as to inspire and unite its
participants under the following envisioned `cause': to bring
together and scrutinize certain important current quantum gravity
research approaches in a fresh, unconventional, almost unorthodox,
way.

\vskip 0.1in

\noindent In part II, a synopsis-{\it cum}-update of work in the
past half-decade or so on applying the algebraico-categorical
concepts, technology and general philosophy of Abstract
Differential Geometry (ADG) to various issues in current classical
and quantum gravity research is presented. The exposition is
mainly discursive, with conceptual, interpretational and
philosophical matters emphasized throughout, while their formal
technical-mathematical underpinnings have been left to the
original papers. The general position is assumed that Quantum
Gravity is in need of a new mathematical, novel physical concepts
and principles introducing, framework in which old and current
problems can be reformulated, readdressed and potentially
retackled afresh. It is suggested that ADG can qualify as such a
theoretical framework.

\vskip 0.1in

\noindent The two parts are closely entwined, as part I makes
general motivating remarks for part II. There are no references in
part I, but part II has numerous citations.

\end{abstract}

\newpage


\newpage

\setlength{\textwidth}{15.9cm} 
\setlength{\oddsidemargin}{0cm}  
\setlength{\evensidemargin}{0cm} 
\setlength{\topskip}{0pt}  
\setlength{\textheight}{21.8cm} 
\setlength{\footskip}{-2cm}

\setlength{\topmargin}{0pt}

\section*{\underline{\LARGE PART I}}

\section*{I.0. Introduction}

Dear participants, on behalf of Professor Anastasios Mallios, the
Algebra and Geometry Section of the Mathematics Department of the
University of Athens, the European Commission (principal sponsors)
and Qualco (private partial sponsors), I wish to welcome you to
the 1st {\itshape Glafka--2004: `Iconoclastic' Approaches to
Quantum Gravity} theoretical physics conference.

\paragraph{An `iconoclast' according to the lexicon.} According to {\itshape Webster's Encyclopedic Unabridged Dictionary of the
English Language}, an `{\em iconoclast}' ({\sl I kon$^{'}$a
klast$^{'}$}, noun) is:

\begin{enumerate}

\item a breaker or destroyer of images, especially those set up
for religious veneration, and/or

\item one who attacks cherished beliefs, traditional institutions,
{\it etc.}, as being based on error or superstition.

\end{enumerate}

\noindent Historically, in Byzantium (723-843AC), `{\em
iconoclasm}' ({\it alias}, `iconomachy') was the polemic movement
against `{\em iconolatry}'---the worshipping of Christian icons
(predominantly in churches).\footnote{In retrospect, I think I
personally would have taken sides with the iconolatres instead of
the iconoclasts after having visited the beautiful Byzantine
Period section of the Benaki National Heritage museum last night.}

The three scientists from past times that immediately spring to
mind as `scientific iconoclasts' are Galileo Galilei, Charles
Darwin and Albert Einstein. The latter revolutionized our ideas of
space, time, matter, energy, and their dynamical
intertransmutations. In view of some challenges presented by
Quantum Gravity (QG), we may have to further revolutionize
Einstein's ideas and thus further `{\em dissect the iconoclast}'.

\paragraph{The twilight of the Quantum Gravity idol.} What is
`{\em the icon}' in our case?: {\em Quantum Gravity}
(QG)---arguably, the `Holy Grail' of theoretical physics in the
dawn of the new millennium. However, there is no quantum theory of
gravity to begin with---anyway, not a conceptually sound,
mathematically consistent and `calculationally' finite one. In a
nutshell, {\em there is no QG icon to destroy in the first place!}
Hence, is our gathering here today `futile', actually `begging the
question' and, ultimately, `begging the quest' for the icon?

Certainly, however, there is a plethora of views and approaches to
QG, so that a `mosaic', `patchwork' sort of picture of QG (with
glaringly conflicting ideas at times!) has emerged over the last
30+ years of research, but there is no unanimous agreement on what
QG is, or anyway, what it ought to be. By the way, theoretical
physicists, unlike religious thinkers and preachers, are
particularly bad when talking about `teleological' and `normative'
aspects of their science, and that's a good thing in my opinion,
as it reflects that they are, in a Socratic sense, not
sure/certain about their knowledge---they have no rigid
convictions that they cannot readily revise or even shed. In
scientific research, uncertainty about a subject is a virtue, not
a blemish. It is sort of liberating not to know, for it invites a
wandering imagination and a way of looking at the World
afresh.\footnote{See the prologue and epilogue of part II.}

\section*{I.1. `First-Order' Iconoclasm}

Thus in our case, `iconoclasm'---at least what I call here `{\em
1st-order iconoclasm}'---pertains to challenging standard or well
established conceptions about and approaches to QG, as well as
proposing alternative ones that are not `mainstream' or
`fashionable' as it were.

The way I see it, the pentaptych of (not mutually independent)
qualities of the theoretical physics' iconoclast are the following
(not in order of import or importance to her research endeavors
and quests):

\begin{enumerate}

\item {\em Imagination} (contra knowledge; ``{\em Imagination is more
important than knowledge}'' (Einstein)---the Glafka
motto\footnote{See Glafka poster.}),

\item {\em `Riskability'} ({\it ie}, able to take risks: `nothing
ventured, nothing gained'---one of Chris Isham's favorite sayings.
Also Wolfgang Pauli: ``{\em Only he who risks has a chance of
succeeding}'',\footnote{See also Richard Feynman's quotation
below.}

\item {\em Obstinacy, perseverance} and {\em `pigheadedness'} (``{\em what
do you care what other people think?}''---Feynman),

\item {\em `Fearlessness'} (especially with regard to making mistakes and
putting one's ideas to the theoretical test and criticism;
Anastasios Mallios).

\item {\em `Authoritilessness'} (question fairly well established
ideas, concepts and practices---take nothing for granted, as a
necessary given; see Einstein quotation below).

\end{enumerate}

\noindent Feynman's words about QG research below, taken from his
Nobel Prize address, epitomize the second virtue of `iconoclasm' I
wanted to highlight for you today:

\begin{quotation}
\noindent ``...{\small It is important that we don't all follow
the same fashion. We must increase the amount of variety and the
only way to do this is to implore you few guys, to take a risk
with your own lives so that you will never be heard of again, and
go off to the wild blue yonder to see if you can figure it
out...}''
\end{quotation}

\noindent Einstein's words bring out the fifth virtue of
`iconoclasm' I wanted to highlight for you today:

\begin{quotation}
\noindent ``...{\small Concepts which have proved useful for
ordering things easily assume so great an authority over us, that
we forget their terrestrial origin and accept them as unalterable
facts. They then become labelled as `conceptual necessities', `a
priori situations', etc. The road of scientific progress is
frequently blocked for long periods by such errors. It is
therefore not just an idle game to exercise our ability to analyze
familiar concepts, and to demonstrate the conditions on which
their justification and usefulness depend, and the way in which
these developed, little by little...}''
\end{quotation}

\noindent While, about obstinacy, perseverance and stubbornly
focusing on a goal, Einstein told once Ernst Strauss:

\begin{quotation}
\noindent ``{\small I know quite certainly that I myself have no
special talent. Curiosity, obsession, and dogged endurance
combined with self-criticism, have brought me to my ideas.
Especially strong thinking power I do not have, or only to a
modest degree. Many have far more of this than I without producing
anything surprising...}''
\end{quotation}

\noindent In this respect, Ernst Straus also relates the following
anecdote about Albert Einstein that I think you will find at least
amusing:

\begin{quotation}
\noindent ``{\small We had finished the preparation of a paper and
we were looking for a paper clip. After opening a lot of drawers
we finally found one which turned out to be too badly bent for
use. So we were looking for a tool to straighten it. Opening a lot
more drawers we came on a box of unused paper clips, Einstein
immediately starting to shape one of them into a tool to
straighten the bent one. When I asked him what he was doing, he
said:} `{\em When I am set on a goal, it becomes very difficult to
deflect me'.}''
\end{quotation}

At the same time, I think it is important that the iconoclast does
not forget that she is {\em standing on the shoulders of giants}
(Isaac Newton); albeit, at the same time standing on her own two
feet...which brings me to what I think of as the `{\em 2nd-order
iconoclasm}'.

\section*{I.2. `2nd-Order' Iconoclasm}

Iconoclasts gather together to tear down each other's
icons---their theories and general `{\it Weltaufbau und
Weltanschaung}', like we have gathered here today. Of course, the
idea is to pick up each other's pieces and synthesize {\em the} QG
icon. For,

\vskip 0.1in

\centerline{\em iconoclasts should not just be `pure
deconstructionists'.}

\vskip 0.05in

\noindent One feels that we ought to find common grounds---as it
were, a common denominator---in our apparently diverse, but
supposedly fundamental and unifying, conceptions of Nature's
depths. We should all have faith in the unity of Physis---after
all, we refer to the World as a Cosmos/$K\acute{o}\sigma\mu
o\varsigma$, not a Chaos/$X\acute{\alpha}o\varsigma$---but we
should also respect and appreciate each other's differences. As
John Archibald Wheeler said: ``{\em More is different}''. We
should search for unity in Nature's cherished
diversity\footnote{Again, see the prologue and epilogue of part
II.}...which brings me to the most radical iconoclasm of the `3rd
kind'.

\section*{I.3. `Third-Order' Iconoclasm}

Here is the paradoxical question:\footnote{In analogy to the
logical oxymoron: `{\em Who shaves the barber?'}.}

\vskip 0.1in

\centerline{\em Who cuts the QG iconoclast?\footnote{In Greek, an
`{\em iconoclast}' (:`$\epsilon\iota\kappa o\nu
o\kappa\lambda\acute{\alpha}\sigma\tau\eta\varsigma$') is (s)he
who `{\em cuts icons}'
(:`$\kappa\lambda\acute{\alpha}\zeta\epsilon\iota~\epsilon\iota\kappa
\acute{o}\nu\epsilon\varsigma$').}}

\vskip 0.05in

\noindent Of course, it is important that my gross {\em
idealization} of the 1st and 2nd-order QG iconoclast
above---especially in view of a not `well defined', let alone
unanimously agreed on, project for a QG theory-construction---does
not turn into an {\em idolization}; for ideally,

\vskip 0.05in

\centerline{\em a genuine iconoclast should tear down all idols,
including (and especially!) his own}.

Thus, to pay my respects to the possibility that {\em we might be
chasing a QG chimera after all}, here is a telling quotation of
David Finkelstein---from an early (:May 1993) pre-print version of
his 1996 book `{\itshape Quantum Relativity: A Synthesis of the
Ideas of Einstein and Heisenberg}' (Springer-Verlag,
1996)---capturing what I coin the (most `radical') {\em
`3rd-order' iconoclasm} of the elusive QG theory {\em itself}:

\vskip 0.1in

\centerline{\underline{\bf The Saviors of Physical
Law}\footnote{This, in a metaphorical sense, `post-anticipates'
Nikos Kazantzakis' `{\em Salvatores Dei}', excerpts of which we
shall encounter in the sequel.}}

\begin{quotation}
\noindent ``...{\small\em What are we after as physicists? Once I
would have said, the laws of nature; then, the law of nature. Now
I wonder.}\footnote{Our emphasis.}

{\small A law, or to speak more comprehensively, a theory, in the
ordinary sense of the word, even a quantum theory of the kind
studied today by almost all quantum physicists, is itself not a
quantum object. We are supposed to be able to know the theory
completely, even if it is a theory about quanta. Its symbols and
rules of inference are supposed to be essentially non-quantum. For
example, ordinary quantum theory assumes that we can know the form
of the equations obeyed by by quantum variables exactly, even
though we cannot know all the variables exactly. This is
considered consistent with the indeterminacies of quantum theory,
because the theory itself is assumed to sum up conclusions from
arbitrarily many experiments.

Nevertheless, since we expect that all is quantum, we cannot
consistently expect such a theory to exist except as an
approximation to a more quantum conception of a theory. At present
we have non-quantum theories of quantum entities. Ultimately the
theory too must reveal its variable nature. For example, the
notion that an experiment can be repeated infinitely often is as
implausible as the notion that it can be done infinitely quickly
($c=\infty$), or infinitely gently ($\hbar=0$).

It is common to include in the Hamiltonian of (say) an electron a
magnetic field that is treated as a non-quantum constant,
expressing the action of electric currents in a coil that is not
part of the endosystem but the exosystem. Such fields are called
external fields. Upon closer inspection, it is understood, the
external field resolves into a host of couplings between the
original electron and those in the coil system, now part of the
endosystem.

{\em It seems likely that the entire Hamiltonian ultimately has
the same status that we already give the external field. No
element of it can resist resolution into further quantum
variables. In pre-quantum physics the ideal of a final theory is
closely connected with that of a final observer, who sees
everything and does nothing. The ideal of a final theory seems
absurd in a theory that has no final observer. When we renounce
the ideal of a theory as a non-quantum object, what remains is a
theory that is itself a quantum object. Indeed, from an
experimental point of view, the usual equations that define a
theory have no meaning by themselves, but only as
information-storing elements of a larger system of users, as much
part of the human race as our chromosomes, but responding more
quickly to the environment. The fully quantum theory lies
somewhere within the theorizing activity of the human race itself,
or the subspecies of physicists, regarded as a quantum system. If
this is indeed a quantum entity, then the goal of knowing it
completely is a Cartesian fantasy, and at a certain stage in our
development we will cease to be law-seekers and become law-makers.

It is not clear what happens to the concept of a correct theory
when we abandon the notion that it is a faithful picture of
nature. Presumably, just as the theory is an aspect of our
collective life, its truth is an aspect of the quality of our
life}\footnote{Again, our emphasis throughout.}...}''
\end{quotation}

\noindent The {\em `law-making'}, as opposed to the (merely) {\em
`law-seeking'}, imperative (of what is here coined `3rd-order
iconoclasm') in the Finkelstein quotation above recalls Nikos
Kazantzakis' concluding words---as it were, the distillation and
r\'esum\'e of his spiritual credo---in his `swan-song' of a book
{\itshape `Salvatores Dei (The Saviours of God): Spiritual
Exercises'}:\footnote{Translated by Kimon Friar (a Touchstone
Book, Simon \& Schuster Publishers, 1960).}

\begin{quotation}
``{\small ...1. Blessed be all those who hear and rush to free
you, Lord, and who say: {\em `Only You and I exist.'}

\vskip 0.1in

2. Blessed too be all those who free you and become united with
you, Lord, and who say: {\em `You and I are One.'}

\vskip 0.1in

3. And thrice blessed be those who bear on their shoulders and do
not buckle under this great, sublime, and terrifying secret: {\em
`That even this One does not exist'}...}''
\end{quotation}

\noindent And with these `agnostic' (but not necessarily
pessimistic!) and `mystical' remarks, I wish you all
wholeheartedly:

\vskip 0.2in

\centerline{\underline{\large\bf Enjoy a mystifying Glafka!}}

\vskip 0.2in

\section*{I.4. Hegelian Postscript: The Owl of Minerva}

And when you thought it was all over, I would like to close this
opening talk with a `{\em post-anticipation}' of the deeper
significance of Glafka, inspired by a recent e-mail exchange with
Rafael Sorkin.

First, I would like to quote Peter Singer---the famous
`bioethicist', from his Princeton homepage:\footnote{http://www.
petersingerlinks.com/minerva.htm}

\begin{quotation}
\noindent ``{\small ...Minerva, the Roman goddess of wisdom, was
the equivalent of the Greek goddess Athena.\footnote{The patron
goddess of Athens.} She was associated with the owl, traditionally
regarded as wise, and hence a metaphor for {\em philosophy}. Hegel
wrote, in the preface to his {\itshape Philosophy of Right}: `{\em
The owl of Minerva spreads its wings only with the falling of the
dusk}'. He meant that philosophy understands reality only after
the event. It cannot prescribe how the world ought to be...}''
\end{quotation}

\noindent Rafael shared with me Balachandran's (:his celebrated
colleague-physicist at Syracuse University) interpretation of
Hegel's owl (which I personally prefer to Singer's strictly `{\em
after-the-fact}' one), according to which:

\begin{quotation}
\noindent ``{\small ...Minerva's owl is spreading its wings at
dusk (or something to that effect), the meaning reputedly being
that only when an event or development is near its end does its
significance become clear...}''
\end{quotation}

\noindent Regarding our Glafka gathering here, it's good that
there's still another 3 days, plus 10 hours or so, till dusk falls
on the last day of the meeting...

\vskip 0.2in

\centerline{$<\bullet>$~$<\bullet>$~$<\bullet>$~$<\bullet>$~$<\bullet>$~$<\bullet>$~$<\bullet>$~$<\bullet>$}

\vskip 0.1in

\section*{\underline{\LARGE \bf PART II}}

\section*{II.0. Prologue: General Motivational Remarks}

Quantum Gravity (QG) has as many facets as there are approaches to
it. There is no unanimous agreement on what QG `really' is---what
are its central questions, its main aims, its basic problems, or
what ought to be ultimately resolved; hence the current `zoo' of
approaches to it. There certainly is overlap between the concepts,
the mathematical techniques and the basic aims of the various
approaches, but the very fact that there are so many different
routes to such a supposedly fundamental quest betrays more our
ignorance rather than our resourcefulness about what QG `truly'
stands for, or at least about how it should be `properly'
addressed and approached.

{\it Prima facie}, the danger that goes hand in hand with the said
proliferation of approaches to QG observed lately is that the {\it
aufbau} of such a theory may eventually degenerate into the
erection of some kind of Babel Tower, where workers working on
each individual approach, just by virtue of the big number of
different, simultaneously developing, schemes (with the
concomitant development of `idiosyncratic' conceptual and
technical jargon, as well as approach-specific mathematical
techniques), may find it difficult to communicate with each other.
As a result, like the mutually isolated seagull populations of the
Galapagos islands that Charles Darwin came across, the various
approaches may eventually cease to be able to cross-breed and the
workers will become `alienated' from each other---{\it ie}, they
will not be able to communicate, let alone to fruitfully interact,
check or cross-fertilize each other's ideas and results. Thus, the
QG vision shall inevitably become disorientated and fragmented;
and what's worse, perhaps irreversibly so. It will then be hard to
believe that all these different workers and their ventures do
indeed have a common goal (:QG), even if they nominally say so
({\it eg}, in conferences!).

Of course, there is that general feeling, ever since the inception
and advent of General Relativity (GR) and subsequently of Quantum
Mechanics (QM), that QG ought to be a coherent amalgamation of
those two pillar theories of 20th century theoretical physics.
Perhaps one of the two theories (or even both!) may have to
undergo significant modifications in order for QG to emerge as a
consistent `unison-by-alteration' of the two. On the other hand,
the gut feeling of many (if not of most) workers in the field is
that, no matter how advanced and sophisticated our technical
(:mathematical) machinery is, we lack the proper
conceptual-physical questions that will open the Pandora's box of
QG. It may well be that the fancy maths get in the way of the
simple fundamental questions we need to come up with in order to
crack the QG `code'. We may be rushing, primarily dazed by past
successes of our mathematical panoply, to give intricate and
complex mathematical answers to simple, yet profound, physical
questions that have not been well posed, or even asked(!), yet.
Fittingly here, Woody Allen's

\vskip 0.1in

\centerline{\small ``I have an answer, can somebody please tell me
the question?''}

\vskip 0.1in

\noindent springs to mind. Time and again the history of the
development of theoretical physics has taught us that in the end,
{\em Nature invariably outsmarts our maths} no matter how
sophisticated and clever they may be, while our own knowledge is
not only insignificant compared to Her wisdom, but also many times
it sabotages the very path that we are trying to pave towards the
fundamental physical questions. For, very often, (mathematical)
knowledge inhibits (physical) intuition and imagination.

Or perhaps, in a promethean sense opposite to that above, it may
be that

\begin{quotation}
\noindent {\em we are not adventurous and `iconoclastic' enough in
our theory-making enterprizes as well as in the mathematical means
that we employ so as to take the `necessary' risks to look at the
QG problem afresh}\footnote{See part I.}---{\it eg}, by creating
new theoretical concepts, new mathematical tools and techniques,
as well as a novel way of philosophizing about them.
\end{quotation}

\noindent In keeping with the `zoological' metaphor above,

\begin{quotation}
\noindent so far the attempts to bring together GR and QM to a
cogent ({\it ie}, a conceptually sound, mathematically consistent,
as well as calculationally finite) QG, seem to this author to be
like {\em trying to cross a parrot with a hyena: so that it (:QG)
can tell us what it is laughing about}.
\end{quotation}

\noindent All in all, it may well be the case that the QG riddle
has been with us for well over half a century now, stubbornly
resisting (re)solution and embarrassingly eluding all our
sophisticated mathematical means of description, because we insist
on applying and trying to marry the `old' physical concepts and
maths---which, let it be appreciated here, have proven to be of
great import in formulating separately the ever so successful and
experimentally vindicated GR and QM---to the virtually unknown
realm of QG.\footnote{This `palindromic' thesis between {\em too
much} and {\em not enough} maths for QG, simply reflects the mean,
neutral position of ignorance, ambivalence and uncertainty of this
author about these matters. See concluding section.} The following
`words of caution' by Albert Einstein \cite{einst7} are very
pertinent to this discussion:

\begin{quotation}
\noindent ``{\small ...Concepts which have proven useful for
ordering things easily assume so great an authority over us, that
we forget their terrestrial origin and accept them as unalterable
facts. They then become labelled as `conceptual necessities', `a
priori situations', etc. The road of scientific progress is
frequently blocked for long periods by such errors. It is
therefore not just an idle game to exercise our ability to analyse
familiar concepts, and to demonstrate the conditions on which
their justification and usefulness depend, and the way in which
these developed, little by little...}'' (1916)
\end{quotation}

In the present paper we take sides more with the second
alternative above, namely, that a new theoretical/mathematical
framework---one that comes equipped with new concepts and
principles, and it is thus potentially able to cast new light on
old ones, as well as to generate new physical questions---is
needed to readdress, reformulate and possibly retackle {\em
afresh} certain caustic, persistently problematic issues in
current QG research. The framework we have in mind is Mallios'
purely algebraico-categorical (:sheaf-theoretic) Abstract
Differential Geometry (ADG) \cite{mall1,mall2,mall4}, while the
account that follows is a semantic, conceptual and philosophical
distillation-{\it cum}-update of results (and their related
aftermath) of a series of applications of ADG to
gravity\footnote{In the sequel, gravity (classical or quantum),
formulated ADG-theoretically, will be coined `{\em ADG-gravity}'
\cite{rap5,rap7,malrap4}.} in the past half-decade or so
\cite{mall2,mall3,malros1,malros2,malros3,malrap1,malrap2,malrap3,malrap4,rap5,rap7,mall9,mall7,mall11,mall10}.
Further details about formal-technical (:mathematical) terms and
results are left to those original papers.

After this introduction, the paper unfolds in three sections, as
follows: in the next section we give a brief {\it r\'esum\'e} of
the principal didactics, as well as the basic physical concepts,
semantics and hermeneutics of ADG. The section that follows it
addresses certain important current classical and quantum gravity
issues under the prism of the background spacetime manifoldless
ADG, and it ends with a brief discussion of current and near
future developments of the theory along topos and more general
category-theoretic lines. The paper closes by continuing the way
it started; {\it ie}, by making general remarks on the
significance and import of a new mathematical-theoretical
framework (such as ADG) in current and future QG research.

\section{II.1. The Basic Tenets and Didactics of ADG}

ADG, we have learned both from theory and from numerous
applications, is a way of doing differential geometry {\em purely
algebraically} (:sheaf-theoretically), without using any notion of
smoothness in the usual sense of Classical Differential Geometry
(CDG)\footnote{In the sequel, the names Differential Calculus (or
simply Calculus) and Analysis shall be regarded as synonyms to the
CDG of smooth manifolds.}---{\it ie}, without employing a base
geometrical differential manifold. {\it In summa}, ADG is a
Calculus-free, entirely algebraic, background manifoldless
theoretical framework of differential geometry
\cite{mall1,mall2,mall4}.

At the basis of ADG lies the notion of $\mathbf{K}$-{\em
algebraized space} ($\mathbf{K}=\mathbf{R},\mathbf{C}$), by which
one means an in principle arbitrary base topological space $X$,
carrying a sheaf $\struc$ of (commutative) $\mathbb{K}$-algebras
($\mathbb{K}=\mathbb{R},\mathbb{C}$) called the {\em structure
sheaf of generalized arithmetics or coordinates}. A family
$\gauge$ of open subsets $U$ of $X$ covering it is called a {\em
system of local open gauges}, while our generalized local
measurements (of coordinates) relative to $\gauge$ are modelled
after the local sections of $\struc$,
$\struc(U)\equiv\Gamma(\gauge\ni U,\struc)$. With $\struc$ in
hand, a {\em vector sheaf} $\modl$ of rank $n$ is a sheaf of
vector spaces of dimensionality $n$ that is locally expressible as
a finite power (:Whitney sum) of $\struc$:
$\modl(U)\simeq\struc^{n}(U)$. By a {\em local gauge frame}
$e^{U}$ ($\gauge\ni U\subset X$), one means an $n$-tuple
$(e_{1},e_{2}\ldots e_{n})$ of local sections of $\modl$ providing
a basis for the vector spaces inhabiting its stalks. Let it be
stressed here that the role of $X$ is just as a `surrogate
scaffolding', which serves as a substrate for the sheaf-theoretic
localization of the objects living in the stalks of the vector and
algebra sheaves involved. $X$ has no physical significance, as we
shall argue below.

One realizes from the beginning how important $\struc$ is in the
theory. We take it almost axiomatically that

\begin{quotation}
\noindent {\em there is no `geometry' without measurement, and no
measurement without a difference}---{\it ie}, what we measure is
always differences or changes in some `measurable' quantities
({\it eg}, coordinates),\footnote{{\it En passant}, let it be
stressed here that it is {\em we} the theorists that declare and
determine up-front what is measurable when we build up our
theories. In this sense, theory and observation are closely tied
to each other (in Greek, `{\em theory}', {\it viz.},
`$\theta\epsilon\omega\rho\acute{\iota}\alpha$', means `a way of
{\em looking} at things'). In a deep sense, we see what {\em we}
want to look at (even in the mind's eye). This also recalls
Einstein's advice to Heisenberg that, apart from the fact that a
theory cannot be built solely on observable quantities, ``{\em it
is the theory that determines what can be observed, not the other
way round}'' \cite{heisenberg}. {\it In toto}, `geometry' is a
creature of the theorist, since it is effectively a mathematical
encodement of and sums up all her observations (:`measurements').
However, as Einstein advised above, in a physical theory not all
entities are `geometrical' (:`observable' or `measurable'). (See
remarks in the sequel about the principal notion of connection
$\conn$ in ADG and ADG-gravity.)} the variability of which is
being secured in our scheme by the fact that, in the case of
coordinates, $\struc$ is a {\em sheaf}.
\end{quotation}

\noindent Indeed, the notion of sheaf is intimately entwined with
that of localization, which physically may be thought of as the
{\em act of gauging physical quantities}, which in turn
essentially promotes them to (dynamically) variable entities. The
bottom-line of all this is that

\begin{quotation}
\noindent the algebras in $\struc$ are {\em differential}
algebras---{\it ie}, they are able to provide us with some kind of
differential operator, via which then we represent the said
(dynamical) changes (:differences).
\end{quotation}

\noindent In turn, we assume that all the `observables'
(:measurable dynamically variable physical quantities) in our
theory can always be expressed in terms of $\struc$ ({\it eg}, as
$\otimes_{\struc}$-tensors).\footnote{$\otimes_{\struc}$ is the
homological tensor product functor.} In a subtle sense,

\begin{quotation}
\noindent from the ADG-theoretic perspective {\em all differential
geometry boils down to the $\struc$ that we choose to use up-front
in the theory's aufbau}.
\end{quotation}

\noindent Parenthetically, but in the same line of thought, we
would like to answer briefly to Shing-Shen Chern's philosophical
pondering in \cite{chern}:

\begin{quotation}
\noindent ``{\small ...A mystery is the role of differentiation.
The analytic method is most effective when the functions involved
are smooth. Hence I wish to quote a philosophical question posed
by Clifford Taubes:\footnote{The reference given here is
\cite{taubes}.} Do humans really take derivatives? Can they tell
the difference?}...''
\end{quotation}

\vskip 0.1in

\noindent by holding that {\em humans do indeed differentiate}
(and they can `really' tell the difference!) {\em insofar as they
can measure}.\footnote{To be precise, in \cite{taubes} Taubes was
talking about so-called {\em inequivalent differential structures}
that a manifold can admit ({\it eg}, {\it \`a la} John Milnor). In
anticipation of the basic ADG-didactics that follow below, our
reply here has a slightly different sense, pertaining to Chern's
mentioning that the most effective method (of differentiating) is
that of Analysis, via smooth manifolds.} From the ADG-theoretic
vantage, they can indeed assume different $\struc$s, provided of
course these structure sheaves of generalized arithmetics
(:coordinates or measurements) furnish them with a differential
operator ({\it viz}. connection) $\partial$. This discussion
brings us to the central notion of ADG.

The neuralgic concept of ADG, as befits any scheme that aspires to
qualify as a theory of {\em differential} geometry proper, is that
of {\em connection} $\conn$ ({\it alias}, generalized differential
$\partial$). $\partial$ (or $\conn$) is categorically defined as a
$\mathbf{K}$-linear, Leibnizian {\em sheaf morphism} between
$\struc$ (or $\modl$), and a sheaf $\Omg$ of $\struc$-modules of
differential form-like entities being the ADG-analogues of the
smooth differential forms encountered in CDG. The connections in
ADG are fittingly coined $\struc$-connections, since $\struc$ is
the `source' of the differential operator $\partial$ (or
equivalently, $\modl\simeq_{\mathrm{loc}}\struc^{n}$ is the
`domain' of $\conn$). In turn, by a {\em field} in ADG, one refers
to the pair $(\modl ,\conn)$, where $\modl$ is the carrier space
of the connection $\conn$.\footnote{This definition of a field may
be thought of as an abstraction and generalization of Yuri Manin's
definition of an electromagnetic (:Maxwell) field as a connection
on a line bundle (although in ADG we do not work with fiber
bundles, but with sheaves, which are more `flexible' and versatile
structures).} The ADG-conception of $\partial$ and $\conn$ is a
Leibnizian ({\it ie}, relational, algebraic), not a Newtonian,
one. That is, in ADG we obtain the differential (structure) from
the algebraic relations (:structure) of the objects living in the
stalks of the vector and algebra sheaves involved, and not from a
background geometrical `space(time)' continuum (:manifold), which
`cartesianly' mediates in our Calculus (ultimately, in our
differential geometric calculations) in the guise of (smooth)
coordinates as in the usual CDG of manifolds.

With $\partial$ and $\conn$ in hand, we can then define the
important notion of {\em curvature} $\curv$ of a connection
$\conn$, an $\struc$-metric $\rho$, torsion, and all the standard
concepts and constructions of the (pseudo-)Riemannian geometry of
GR; albeit, to stress it again, entirely algebraico-categorically,
without using any background geometrical locally Euclidean
(:manifold) space(time). $\curv$, like $\conn$, is a sheaf
morphism, but unlike its underlying connection which is only a
$\mathbf{K}$-morphism, it is an $\struc$-morphism (or
$\otimes_{\struc}$-tensor). The dynamical relations (:physical
laws) between the observable physical quantities noted above are
then expressed differential geometrically as differential
equations proper. In other words,

\begin{quotation}
\noindent in ADG the laws of physics are categorically expressed
as {\em equations between sheaf morphisms},\footnote{From this it
follows what we noted earlier, namely, that the base arbitrary
topological space $X$ plays absolutely no role in the physical
dynamics in our theory.} such as the curvature of the connection.
\end{quotation}

\noindent In ADG-gravity in particular, the vacuum Einstein
equations are formulated in terms of the Ricci scalar curvature
$\ricci$ of a gravitational connection $\conn$:\footnote{This is
the only displayed mathematical expression in the present paper!}

\begin{equation}\label{eq1}
\ricci(\modl)=0
\end{equation}

\noindent Perhaps the deepest observation one can make about
(\ref{eq1}) above is that it is an `$\struc$-{\em functorial}'
expression. This means that the Einstein equations are expressed
via the curvature of the connection (and not directly in terms of
the connection itself!), which as noted above is an
$\struc$-morphism (:an $\otimes_{\struc}$-tensor). The
gravitational field, in the guise of $\curv(\conn)$, `sees
through' and it is unaffected ({\it ie}, it remains `invariant')
by our generalized measurements in $\struc$. This is a {\em
categorical} description of the ADG-analogue of the Principle of
General Covariance (PGC) of GR, which group-theoretically may be
represented by $\aut\modl$ as we shall note in the next section.
In connection with the discussion around footnote 18 above, it is
interesting to note that the principal entity in ADG-gravity, the
gravitational connection $\conn$, strictly speaking is {\em not}
itself an `observable'---{\it ie}, a {\em measurable} dynamical
entity in the theory---as it is {\em not} a `geometrical object'
(:an $\struc$-morphism or $\otimes_{\struc}$-tensor). However, its
curvature $\curv(\conn)$ is an observable, and the vacuum Einstein
equations (\ref{eq1}) are expressed via it.\footnote{On this
remark hinges the observation that {\em $\conn$ is not a
geometrical entity}; rather, it is an {\em algebraic} (:analytic)
one. (See also Anastasios Mallios' contribution to this volume
\cite{mall10}.)} The moral here {\it vis-\`a-vis} Einstein's
advice to Heisenberg in footnote 5, is that the central notion in
ADG-gravity (and in ADG in general)---that of connection
$\conn$---is an `unobservable' entity, as it eludes our
generalized coordinates (:measurements) in $\struc$.

In turn, on the last observation above rests our generalized
Principle of Field Realism (PFR), which is closely related to our
categorical version of the PGC of GR noted earlier
(:$\struc$-functoriality), and roughly it maintains that

\begin{quotation}
\noindent The ADG-gravitational field $\conn$, and the field law
(\ref{eq1}) that it defines differential geometrically (:as a
differential equation proper), remains unaffected (and the
corresponding law `invariant') by our `subjective', arbitrary
choices of $\struc$.
\end{quotation}

\noindent Einstein's words below, taken from his `{\itshape Time,
Space, and Gravitation}' article in \cite{einst10} where he gives
an account of how he arrived at the PGC of GR as `invariance of
the law of gravity under arbitrary coordinate transformations',
are very relevant here:

\begin{quotation}
\noindent ``{\small ...Must the independence of physical laws with
regard to a system of coordinates be limited to systems of
coordinates in uniform movement of translation with regard to one
another? {\small\em What has nature to do with the coordinate
systems that \underline{we} propose and with their
motions?}\footnote{Our emphasis.} Although it may be necessary for
our descriptions of nature to employ systems of coordinates that
we have selected arbitrarily, the choice should not be limited in
any way so far as their state of motion is concerned\footnote{Or
perhaps better expressed, (the said arbitrary choice of any
particular system of) coordinates should not affect in any way the
dynamical equations (laws) of motion of the fields in
focus.}...}''
\end{quotation}

\noindent the subtle but important generalization of the PGC of GR
by ADG-gravity culminating in the PFR above is that

\begin{quotation}
\noindent the field law of gravity remains unaffected
(:`invariant') not only by arbitrary (:general) {\em smooth}
coordinate transformations ({\it ie}, by general transformations
of coordinates within the structure sheaf
$\struc\equiv\smooth_{M}$ chosen by the theorist/`observer'), but
also by arbitrary changes of $\struc$ itself.
\end{quotation}

\noindent In our work this last remark has been promoted to a
principle, coined the Principle of Algebraic Relativity of
Differentiability (PARD), and it maintains that

\begin{quotation}
\noindent no matter what $\struc$ is chosen to furnish us with,
and thus to geometrically represent (in $\modl$), the
gravitational field $\conn$, the field law of gravity that the
latter defines remains unaffected by it.
\end{quotation}

\noindent Thus, as a pun to Taubes' question that Chern was quoted
as asking in the previous section, we can now retort: {\em the
ADG-gravitational connection field is indifferent to different
choices of differential algebras of generalized coordinates
$\struc$ that we employ to represent it (on $\modl$)}. For, to
emulate Einstein's words above: {\em what has nature (here, the
gravitational field law) to do with the $\struc$s that we choose
to geometrically represent (via $\modl$) the (inherently
algebraic) gravitational field $\conn$?}

In closing this section, it must be stressed in view of the last
remarks and footnote above that the generalized coordinates in
$\struc$, once they supply us with the differential geometric
mechanism---{\it ie}, with the differential $\partial$ or the
connection $\conn$---they are effectively ({\it ie}, as far as the
expression of the field law of gravity is concerned) `discarded'
as they have absolutely no physical significance, since the
gravitational field dynamics (\ref{eq1}) `sees through' them (:it
is $\struc$-covariant, or $\struc$-functorial). It took Einstein
more than 7 years to appreciate the metric and hence the
dynamical\footnote{Since in GR {\it \`a la} Einstein, the metric
$g_{\mu\nu}$ is the sole dynamical variable.} insignificance of
coordinates; albeit, the smooth base spacetime manifold
(:$\struc_{X}\equiv\smooth_{M}$) is invaluable in standard GR, if
anything, in order to formulate the theory differential
geometrically ({\it ie}, to model the dynamics after differential
equations proper) \cite{kriele}.

{\it In toto}, in GR too, the Einstein equations are generally
covariant since they are formulated as differential equations
between smooth, $\otimes_{\smooth_{M}}$-tensors. The subtle point
here is that in the manifold and CDG-based GR, whenever a concrete
calculation is made, the smooth coordinates are invoked and the
background spacetime continuum provides us with a
geometro-physical interpretation of the theory. That is, in GR,
spacetime events and smooth spacetime intervals between them have
a direct experimental meaning, as they are `quantities' to be
measured (:recall that $g_{\mu\nu}$ represents both the
gravitational field and the spacetime chronogeometry). By
contrast, in the purely algebraic ADG-gravity, there is {\it a
priori} no need for a geometrical (smooth) spacetime
interpretation of the theory.\footnote{This doing away with the
smooth background geometrical spacetime manifold of ADG-gravity
proves to be very important in both classical and quantum gravity
current research as we shall argue in the next section.} Here is a
challenging question for future physical applications of ADG:

\begin{quotation}
\noindent {\em Can we relate the theory (:ADG-gravity) to
experience directly from its purely algebraic underpinnings,
without recourse to a background geometrical manifold
representation and its associated spacetime
interpretation?}\footnote{This author is indebted to the referee
of \cite{rap5} for bringing him to ask this question with his
acute remarks on the connection between ADG-gravity's doing away
with coordinates and experiment.}
\end{quotation}

\section{II.2. Implications of Background Spacetime Manifoldlessness}

In this section we outline the main `aftermaths'---{\it ie}, the
results following the application of the ADG-maths (pun
intended)---of numerous applications of the base spacetime
manifoldless ADG to gravity. To prevent the reader's distraction
from repeated referencing within the text, the citations where all
the results that follow can be found are
\cite{mall2,mall3,malros1,malros2,malros3,malrap1,malrap2,malrap3,malrap4,rap5,rap7,mall9,mall7,mall11,mall10}.

\paragraph{ADG-gravity as pure gauge theory of the 3rd kind.} ADG-gravity
has been called `{\em pure gauge theory of the third kind}' due to
the following three characteristic features:

\begin{itemize}

\item First, the sole dynamical variable in ADG-gravity is the
$\struc$-connection $\conn$. This is in contradistinction to the
original second-order formalism of GR due to Einstein in which the
sole dynamical variable is the spacetime metric $g_{\mu\nu}$ whose
ten components represent the gravitational potentials, or even to
the recent first-order Palatini-type of formalism due to Ashtekar
in which two gravitational variables are involved---the tetrad
field $e_{\mu}$ and the spin-Lorentzian connection
$\aconn$.\footnote{Let it be noted here that the smooth metric of
the original 2nd-order formalism is still present `in disguise' in
Ashtekar's scheme \cite{ash}, as $g_{\mu\nu}$ is effectively
encoded in the {\it vierbein} $e_{\mu}$s.} Fittingly, the
ADG-formulation of gravity has been called `{\em half-order
formalism}', since only half the variables (namely, only the
connection) of the first-order formalism are involved.

\item Second, due to the manifest absence of a background geometrical
smooth spacetime manifold $M$, there is no distinction between
external (:spacetime) and internal (:gauge) symmetries. In
ADG-gravity, the $\diff$ of external smooth spacetime symmetries,
traditionally implementing the PGC in the manifold and, {\it in
extenso}, the CDG-based GR, is replaced by $\aut\modl$---the
principal group sheaf of automorphisms of the ADG-gravitational
field $(\modl ,\conn)$. Of course, by virtue of the local
isomorphism $\modl|_{U}\simeq\struc^{n}$, $\aut\modl$ assumes
locally the more familiar form:
$\aut\modl|_{U}=\mathcal{G}\mathcal{L}(n,\struc(U))$---the group
sheaf of general (generalized) coordinates' transformations. This
is a Kleinian perspective on field geometry: the geometry of the
field (:and concomitantly, of the law that it defines) is its
automorphism group (:and concomitantly, the symmetries of the law
that it defines).

\item And third, from the above it follows that ADG-gravity is neither a
gauge theory of the 1st kind (:global gauge symmetries, global
gauge frames), nor one of the 2nd kind (:spacetime localized gauge
symmetries, local gauge frames). There is no external, to the
ADG-gravitational field $(\modl ,\conn)$, spacetime. The field is
a dynamically autonomous entity, whose `auto-symmetries'
(:`self-invariances' of the law (\ref{eq1}) that it defines) are
encoded in $\aut\modl$. This makes the ADG-gravitational field an
autonomous, `{\em external spacetime unconstrained gauge system}'.
As a result, in ADG-gravity there is no distinction between
external (:`spacetime') and internal (:`gauge') symmetries: all
symmetries are `esoteric' to the field, pure gauge ones.

\end{itemize}

In view of the above, the `{\em background smooth spacetime
manifoldless half-order formalism}' of ADG-gravity may shed light
on the outstanding problem of treating gravity as a gauge theory
proper \cite{ivanenko}---a problem which is largely due to our
persistently fallacious viewing of $\diff$ as a gauge group proper
\cite{weinstein}.

In the absence of an external (:background) geometrical spacetime
manifold $M$ and the autonomous conception of the gravitational
field in ADG-gravity, we encounter no problems originating from
$M$ and its $\diff$ `structure group'. On the other hand, the
classical theory (GR), as well as various attempts to quantize it
by retaining the base $M$ and hence the entire CDG-technology, do
encounter such problems---one of them being the problem of
regarding gravity as a gauge theory proper mentioned above. Let us
discuss some more of them.

\paragraph{The role of singularities in ADG-gravity.} The role of
singularities in GR was well known and appreciated since the times
of Einstein and Schwarzschild, but it got worked out and further
clarified in the celebrated works of Hawking and Penrose in the
late 60s/early 70s. Briefly, singularities are thought of as {\it
loci} in the spacetime continuum where some physically important
quantity grows without bound and, ultimately, the Einstein
gravitational equations seem to break down. Given some generic
conditions, the Einstein equations appear to `predict'
singularities---sites of their own destruction. This is pretty
much the general aftermath of the manifold based Analysis of
spacetime singularities \cite{clarke4}. In this Analysis (and this
is the general consensus in gravitational physics), although
singularities are pushed to the boundary of an otherwise regular
spacetime manifold, they are regarded as being physically
significant, in spite of Einstein's position to the contrary till
the end of his life \cite{einst3}:

\begin{quotation}
\noindent ``{\small ...A field theory is not yet completely
determined by the system of field equations. Should one admit the
appearance of singularities?...{\small\em It is my opinion that
singularities must be excluded. It does not seem reasonable to me
to introduce into a continuum theory points (or lines {\it etc.})
for which the field equations do not hold}\footnote{Our
emphasis.}...}''
\end{quotation}

\noindent In this line of thought however, few would doubt that
the main culprit for the singularities of GR is the smooth base
spacetime manifold which is {\it a priori} assumed in the theory,
in the sense that every singularity is a pathology of a smooth
function in $\smooth_{M}$---the sheaf of germs of smooth functions
on $M$.\footnote{Here it is tacitly assumed that a differential
manifold $M$ is nothing else but the algebra $\smooth(M)$ of
smooth functions on it (Gel'fand duality).} Moreover, the very PGC
of GR, which is mathematically implemented via $\diff$ as noted
before, appears to come in conflict with the existence of
gravitational singularities, which makes a precise definition of
the latter perhaps the most problematic issue in GR
\cite{geroch,clarke4}.

By contrast, in the base spacetime manifoldless ADG-gravity,
singularities are not thought of as breakdown points of the law of
gravity, at least not in any differential geometric sense. Quite
on the contrary, {\em the ADG-formulated Einstein equations are
seen to hold over singularities of any kind}. This is not so much
a `{\em resolution}' of singularities in the usual sense of the
term, as an `{\em absorption}' of them in the ADG-gravitational
field $(\modl ,\conn)$. That is, singularities are incorporated in
$\struc$ (thus, in effect, they are absorbed in $\modl$), in the
sense that they are singularities of some functional, generalized
coordinate-type of entity in the structure sheaf of generalized
arithmetics that {\em we} choose in the first place to employ in
the theory. The aforementioned $\struc$-functoriality of the
ADG-gravitational dynamics secures that the ADG-gravitational
field `sees through' the singularities carried by $\struc$, and
the latter in no sense are breakdown {\it loci} of the
differentially (:differential geometrically) represented field law
of gravity as a differential equation proper as the manifold and
CDG-based analysis of spacetime singularities has hitherto made us
believe \cite{clarke4}. Thus, in view of the ADG-generalized PGC
and its associated PFR mentioned in the previous section,
Einstein's `non-belief' in singularities can be succinctly
justified in ADG-gravity as follows:

\begin{quotation}
\noindent What has nature (here, the physical field of gravity and
the law that it defines as a differential equation) to do with
coordinates (here, $\struc$) and the singularities that they
carry? If coordinates are unphysical because they do not partake
into the ADG-gravitational dynamics (:$\struc$-functoriality of
(\ref{eq1})), then so are singularities, since they are inherent
in $\struc$.
\end{quotation}

Nevertheless, the general opinion nowadays is that, although
gravitational singularities are a problem of classical gravity
(GR) long before its quantization becomes an issue, a quantum
theory of gravity should, if not remove them completely much in
the same way that quantum electrodynamics did away with the
unphysical infinities in Maxwell's theory, at least show us a way
towards their resolution \cite{pen5}. We thus turn to some quantum
implications of the base manifoldless ADG-gravity and how the
singularity-absorption into $\struc$ mentioned above may come in
handy.

\paragraph{Towards a 3rd-quantized theory of gravity.} The
ADG-theoretic outlook on gravity is field-theoretic {\it par
excellence}. In fact, it is purely 3rd-gauge field-theoretic, as
it employs solely the algebraic connection field and there is no
external (to the field) geometrical spacetime manifold.

From a geometric (pre)quantization and 2nd (:field) quantization
vantage, the (local) sections of $\modl$ represent (local) quantum
particle (position) states of the field.\footnote{Indeed, $\modl$
may be thought of as the associated (:representation) sheaf of the
principal group sheaf $\aut\modl$ of field automorphisms.}
Moreover, these `field quanta' obey an ADG-analogue of the
spin-statistics connection: extending to vector sheaves
Selesnick's bundle-theoretic musings in \cite{sel}, boson states
correspond to sections of {\em line} sheaves\footnote{Vector
sheaves of rank $1$.}, while fermions are represented by sections
of vector sheaves of rank greater than $1$.

Parenthetically, it must be noted here that the said
representation of (gauge and matter) particle-quanta states as
sections of the corresponding $\modl$s ties well with the
aforesaid incorporation of singularities in $\struc$ (or $\modl$),
in the following sense: ever since the inception of GR, and
subsequently with the advent of QM, it is well reported that
Einstein in his unitary field theory program\footnote{Which, let
it be noted here, was intended to `explain away' QM altogether.}
wished to describe the particle-quanta as `{\em singularities in
the field}'. Prophetically, Eddington \cite{eddington2}
anticipated him:

\begin{quotation}
\noindent ``{\small ...It is startling to find that the whole of
dynamics of material systems is contained in the law of
gravitation; at first gravitation seems scarcely relevant in much
of our dynamics. But there is a natural explanation. {\small\em A
particle of matter is a singularity in the gravitational
field},\footnote{Our emphasis.} and its mass is the pole-strength
of the singularity; consequently {\small\em the laws of motion of
the singularities must be contained in the
field-equations},\footnote{Again, our emphasis.} just as those of
electromagnetic singularities (electrons) are contained in the
electromagnetic field-equations...}''
\end{quotation}

\noindent By absorbing the singularities into $\struc$, by
identifying quantum-particle states as sections of $\modl$ ({\it
ie}, in effect of $\struc$!), and by the $\struc$-functoriality of
the ADG-gravitational dynamics, we have a direct realization of
Eddington's anticipation above: {\em the particle-quanta co-vary
with the field-law itself}. In a strong de Broglie-Bohmian sense,
the connections are the `{\em guiding fields}' of their particles:
they embody them and carry them along the dynamics (:field
equations) that they define.

The upshot of all this is that, due to the external spacetime
manifoldlessness of the theory, the quantum perspective on
ADG-gravity:

\begin{itemize}

\item May be coined 3rd-quantum field theory.\footnote{Recently, Petros Wallden brought
to the attention of this author that the term `{\em third
quantization}' has already been used in quantum gravity and
quantum cosmology research \cite{strominger}. However, the sense
in which we use this term is quite different from that.} {\it In
toto}, {\em QG from the ADG-perspective is a 3rd-quantum,
3rd-gauge field theory}.

\item Since the ADG-gravitational field is an external spacetime unconstrained gauge system,
there is also {\it prima facie} no problem in defining (gauge
invariant) observables in (vacuum) Einstein gravity \cite{torre1},
or a (physical) inner product (:physical Hilbert space); while no
problem of time arises either, since $\diff$ is absent from the
theory from the very start \cite{ish2,torre2}.\footnote{All these
problems are encountered in the manifold (and CDG) based canonical
approaches to QG, in which the gravitational field is viewed as a
spacetime constrained gauge system and $\diff$ represents those
so-called primary space-time constraints (:in a canonical
$3+1$-split smooth spacetime manifold setting, the primary
constraints are the $3$-spatial diffeos and the Hamiltonian
time-diffeo resulting in the celebrated Wheeler-de Witt equation
satisfied by physical states).}

\item In a possible covariant (:path integral) quantization of ADG-gravity,
the physical configuration space is the moduli space of the affine
space $\sconn$ of $\struc$-connections, modulo the field's gauge
auto-transformations in $\aut\modl$. Here too, since $\diff$ is
not present, there should be no problem in finding a convenient
measure to implement the said functional integral. Towards this
end, and with some new ADG-results in hand \cite{mall4},
Radon-type of measures on $\sconn /\aut\modl$ are currently being
investigated. There have been recent QG tendencies to develop
differential geometric ideas and a related integration theory on
the moduli space of gravitational connections, as for example in
Loop Quantum Gravity (LQG) \cite{ashlew2,ashlew5,smolin}, but
advances appear to be stymied by the ever-present background
smooth spacetime manifold and its associated $\diff$
\cite{baez2,baez3}.

\item {\em There is no quantization of spacetime {\it per se} entertained in
ADG-gravity, since there is no spacetime to begin with}. Such a
spacetime quantization procedure figures prominently in current
gauge-theoretic ({\it ie}, connection based) approaches to QG such
as LQG, and it is used there to resolve smooth spacetime
singularities \cite{modesto,husain}. Thus here we have an instance
of the aforesaid general anticipation of current QG researchers,
namely, that a quantum theory of gravity should remove
singularities. Indeed, LQG appears to resolve singularities via
spacetime quantization. Again, this must be contrasted against
ADG-gravity, where {\it ab initio} there is no spacetime continuum
hence no spacetime quantization either, while singularities are
being absorbed in the field law itself; hence, strictly speaking,
there is no need for their `quantum resolution'.

\item Last but not least comes the issue of the formulation of a
manifestly {\em background independent} non-perturbative QG
\cite{ash5,alvarez,smolin}. Normally, `background independence'
means `{\em background geometry (:metric) independence}'.
ADG-gravity is explicitly background metric independent, since no
metric is involved in the theory ({\it ie}, the aforementioned
$\struc$-metric has no physical significance---it is not a
dynamical variable---in the theory).\footnote{It is an optional,
auxiliary structure externally (to the field $\conn$) imposed by
the experimenter (:`observer' or `measurer'); much like $\struc$
itself.} Furthermore, unlike the current connection based
approaches to QG, which vitally rely on a background smooth
manifold for their differential geometric concepts and
constructions, {\em ADG-gravity is manifestly background spacetime
manifold independent}.

\end{itemize}

Thus, in view of all the virtues of ADG-gravity above, one is
tempted to ask the following couple of questions:

\begin{quotation}

\noindent $\bullet$ In the guise of (\ref{eq1}), {\em don't we
already possess a quantum version of the (vacuum) Einstein
equations?}

\vskip 0.05in

\noindent and concomitantly:

\vskip 0.05in

\noindent $\bullet$ Since not only a background metric, but also a
background spacetime (manifold) is {\em not} involved in the
theory, does the need arise to {\em quantize spacetime itself}?

\end{quotation}

\noindent The immediate reply is `{\em yes}' and `{\em no}',
respectively.

\paragraph{The future in a nutshell: QG in a topos.} The last paragraph
in the present section is concerned with the possibility of
formulating ADG-theoretically QG in a topos. A topos is a special
type of category that can be interpreted both as an abstract
`pointless space' and as a `logical universe of variable
mathematical entities'. In a topos, geometry and logic are unified
\cite{macmo}. Thus, the basic intention here is to organize the
sheaves involved in ADG-gravity into a topos-like structure in
which deep logico-geometrical issues in QG can be addressed. A
mathematical byproduct of such an investigation would be to link
ADG with the topos-theoretic Synthetic Differential Geometry (SDG)
of Kock and Lawvere \cite{kock,laven}, which in turn has enjoyed
various applications so far to classical and quantum gravity
\cite{grink,guts0,guts1,guts2,guts3,guts,buttish4,ish3}. In this
respect, of purely mathematical interest would be to compare and
try to bring together under a topos-theoretic setting the
principal notion of both ADG and SDG---that of {\em connection}
\cite{mall-3,mall1,vas1,kock1,kock,laven}. In the context of a
finitary, causal and quantal version of Lorentzian gravity
formulated in ADG-terms
\cite{malrap1,malrap2,malrap3,rap5,malrap4}, this enterprize (with
a Grothendieck topos twist closely akin to a recent approach to
quantum geometry and QG coined `Causal Site Theory'
\cite{crane}\footnote{A categorical generalization of the `Causal
Set Theory' of Sorkin {\it et al.}
\cite{sorkin1,sork2,sork3,sork4}.}) has already commenced
\cite{rap7}.\footnote{Anticipatory works of such an enterprize are
\cite{rap0,rap4,rap6}.}

Another categorical approach to QG which ADG-gravity could in
principle be related to is the recent `{\em Quantizing on a
Category}' (QC) general mathematical scheme due to Isham
\cite{ish5,ish6,ish7,ish8}. The algebraico-categorical QC is
closely akin to ADG both conceptually and technically, having
affine basic motivations and aims. QC's main goal is to quantize
systems with configuration (or history) spaces consisting of
`points' having internal (algebraic) structure. The main
motivation behind QC is the failure of applying the conventional
quantization concepts and techniques to `systems' ({\it eg},
causets or spacetime topologies) whose configuration (or general
history) spaces are far from being structureless-pointed
differential manifolds. Isham's approach hinges on two
innovations: first it regards the relevant entities as objects in
a category, and then it views the categorical morphisms as
abstract analogues of momentum (derivation maps) in the usual
(manifold based) theories. As it is the case with ADG, although
this approach includes the standard manifold based quantization
techniques, it goes much further by making possible the
quantization of systems whose `state' spaces are not smooth
continua.

Indeed, there appear to be close ties between QC and
ADG-gravity---ties which ought to be looked at closer. {\em Prima
facie}, both schemes concentrate on evading the (pathological)
pointed differential manifold---be it the configuration space of
some classical or quantum physical system, or the background
spacetime arena of classical or quantum (field) physics---and they
both employ `pointless', categorico-algebraic methods. Both focus
on an abstract (categorical) representation of the notion of
derivative or derivation: in QC, Isham abstracts from the usual
continuum based notion of vector field (derivation), to arrive at
the categorical notion of arrow field which is a map that respects
the internal structure of the categorical objects one wishes to
focus on (eg, topological spaces or causets); while in our work,
the notion of derivative is abstracted and generalized to that of
an algebraic connection, defined categorically as a sheaf
morphism, on a sheaf of suitably algebraized structures ({\it eg},
causal sets, or finitary topological spaces and the incidence
algebras thereof representing quantum causal sets, as in the
finitary version of ADG-gravity
\cite{malrap1,malrap2,malrap3,rap5,rap7}).

\section{II.3. Epilogue: General Closing Remarks}

In this epilogue we would first like to discuss whether it is
still reasonable to believe that we can use differential geometric
ideas in the quantum deep, that is, in the QG domain. Then, we
would like to conclude this paper by continuing the general theme
of the prologue, namely, that QG research is in need of new
concepts, new mathematics, and a novel way of philosophizing about
them.

\paragraph{Still use differential geometry in QG?} Although the general feeling
nowadays among theoretical physicists (and in particular, `quantum
gravitists') is that below a so-called Planck length-time
($\ell_{P}$-$t_{P}$),\footnote{$\ell_{P}=\sqrt{\frac{G\hbar}{c^{3}}}=1.6\times
10^{-33}cm;~ t_{P}=\sqrt{\frac{G\hbar}{c^{5}}}=5.3\times
10^{-44}s$.} where quantum gravitational effects are supposed to
become significant, the space-time continuum (:manifold) should
give way to something more reticular (:discrete) and quantal,
CDG-ideas and technology still abound in current QG research.
Consider for instance the manifold based CDG used in all its glory
in the canonical and covariant approaches to QG ({\it eg}, LQG
\cite{ashlew2,ashlew5}), or the higher-dimensional (real analytic
or holomorphic) manifolds ({\it eg}, Riemann surfaces, K\"ahler
manifolds, Calabi-Yau manifolds, supermanifolds, {\it etc.})
engaged in (super)string theory research, or even the so-called
noncommutative differential spaces that Connes' Noncommutative
Differential Geometry propounds \cite{kastler,connes,connes1},
which are still, deep down, differential manifolds in the usual
sense of the term. {\it In toto}, smooth manifolds and CDG are
still well and prosper in QG.

A few people, however, have aired over the years serious doubts
about whether the spacetime continuum and, {\it in extenso}, the
CDG that is based on it, could be applied {\em at all} in the QG
domain. Starting (in chronological order) with Einstein, then
going to Feynman, the doubts reach their climax in Isham's
categorematic `{\em no-go of differential geometry in QG}' below:

\begin{quotation}
\noindent ``{\small `...You have correctly grasped the drawback
that the continuum brings. If the molecular view of matter is the
correct (appropriate) one; {\it ie}, if a part of the universe is
to be represented by a finite number of points, then the continuum
of the present theory contains too great a manifold of
possibilities. I also believe that this `too great' is responsible
for the fact that our present means of description miscarry with
quantum theory. {\em The problem seems to me how one can formulate
statements about a discontinuum without calling upon a continuum
space-time as an aid; the latter should be banned from theory as a
supplementary construction not justified by the essence of the
problem---a construction which corresponds to nothing real. But we
still lack the mathematical structure unfortunately}.\footnote{Our
emphasis.} How much have I already plagued myself in this way [of
the manifold]!...}'' \cite{stachel}
\end{quotation}

\centerline{.............................}

\begin{quotation}
\noindent ``{\small{\em ...The theory that space is continuous is
wrong, because we get...infinities} {\small\rm [ viz.
`singularities']} {\em\small and other similar difficulties}
...{\small\rm [ while]} {\small\em the simple ideas of geometry,
extended down to infinitely small, are wrong\footnote{Our emphasis
throughout.}...}}'' \cite{feyn1}
\end{quotation}

\centerline{.............................}

\begin{quotation}
\noindent ``{\small{\em ...At the Planck-length scale,
differential geometry is simply incompatible with quantum
theory}...{\small [so that]} {\small\em one will not be able to
use differential geometry in the true quantum-gravity
theory\footnote{Our emphasis.}...}}'' \cite{ish}
\end{quotation}

\vskip 0.1in

\noindent Isham's remarks are shrewd, critical and iconoclastic:

\begin{quotation}
{\em CDG and the classical $\smooth$-smooth manifold model of
spacetime supporting its constructions `miscarry with'} (to use
Einstein's expression above) {\em quantum theory, and it will
therefore be of no import to QG research}.
\end{quotation}

\noindent On the other hand, and this is one of the basic
aftermaths of our work, from an ADG-theoretic point of view it is
not exactly that differential geometric ideas cannot be used in
the quantum regime---as if the intrinsic differential geometric
mechanism (which in its essence is of an algebraic nature) fails
in one way or another when applied to the realm of QG---but rather
that when that mechanism is geometrically effectuated or
implemented (represented) by the (cartesian mediation in the guise
of the smooth coordinates of the) background $\smooth$-smooth
spacetime manifold as in CDG, then all the said problems
(:singularities, unphysical infinities, $\diff$-related
pathologies) crop up and are insurmountable (always within the
confines of, {\it ie}, with the concepts and the methods of, the
theoretical framework of the manifold based Analysis).

Thus, to pronounce this subtle but crucial from the
ADG-perspective difference, we maintain that

\begin{quotation}
\noindent the second part of Isham's quotation above should also
carry the adjective `{\em classical}' in front of `{\em
differential geometry}', and read: `{\em one will not be able to
use \underline{classical} differential geometry}' (or
equivalently, a geometrical base differential spacetime manifold)
`{\em in the true quantum-gravity theory}'.
\end{quotation}

\noindent {\it In summa}, the aforesaid subtle distinction hinges
on the physical non-existence of a background geometrical smooth
spacetime manifold, {\em not} of the inapplicability of the {\em
essentially algebraic mechanism of differential geometry}, which
can still be retained and applied to QG research. Metaphorically
speaking, ADG-gravity has shown us a way {\em not} to throw away
the baby (:the invaluable algebraic differential geometric
mechanism) together with the bath-water (:the base smooth
spacetime manifold). The `icon' (or perhaps better, the `idol')
that Isham's iconoclastic words ought to cut out of physics once
and for all is the background geometrical spacetime manifold and
{\em not} the invaluable differential geometric machinery which
CDG has so far misled us into thinking that is inextricably tied
to the base manifold.

To summarize, in the background geometrical spacetime manifoldless
ADG-gravity, all the classical and quantum gravity problems we
mentioned in the previous section, which are all due to the base
$M$, its $\diff$ and, {\it in extenso}, to the CDG that is based
on the latter, simply disappear---{\it ie}, they become
non-problems. Thus, ADG does not solve these puzzles; it simply
cuts the Gordian knot that they present us within the
CDG-framework. This is analogous to how Wittgenstein \cite{witt1}
maintained that philosophical problems could be solved: simply by
changing perspective---ultimately, by changing theoretical
framework:

\begin{quotation}
\noindent ``{\small ...The solution of philosophical problems can
be compared with a gift in a fairy tale: in the magic castle it
appears enchanted and if you look at it outside in daylight it is
nothing but an ordinary bit of iron (or something of the
sort)\footnote{Our emphasis throughout.}...}''
\end{quotation}

\noindent Indeed, problems in GR like that of singularities and
Einstein's hole argument
\cite{stachel3,stachel5,stachel0,stachel9}, as well as the problem
of time and that of observables in QG, look formidable (in fact,
insuperable!) when viewed and tackled via the manifold based
CDG---ultimately, when we are bound by ``{\em the golden shackles
of the manifold}'' \cite{ish}. However, under the light of ADG,
`{\em gold looks nothing but an ordinary bit of iron}'.
Furthermore, much in the same way that Wittgenstein in
\cite{witt2} contended that

\begin{quotation}
\noindent ``{\small ...Our task is, not to discover new calculi,
but to describe the present situation in a new light...}''
\end{quotation}

\noindent our ADG-framework (and, as a result, ADG-gravity), does
not purport to be some kind of new Differential Calculus (and,
accordingly, ADG-gravity a new theory of gravitation); it simply
goes to show that most (if not all!) of the differential geometric
mechanism `inherent' in CDG can be articulated entirely
algebraically, without the cartesian mediation of a background
geometrical (spacetime) manifold (with all the supposedly physical
pathologies that the latter is pregnant to). In addition, it goes
without saying that if the base geometrical $M$ has to go, so must
the geometrical (spacetime) interpretation of the theory
(:GR).\footnote{Of course, it now behooves us to answer to the
question posed at the end of section II.1.}

For after all, Einstein too, overlooking the great success that
the geometrical spacetime manifold based GR enjoyed during his
lifetime, insisted that:

\begin{quotation}
\noindent ``{\small ...Time and space are modes by which {\em we
think}, not conditions in which we live}''
\cite{einst2}\footnote{This quotation can also be found in
Anastasios Mallios' contribution to this volume.}...``{\small [the
spacetime continuum] corresponds to nothing real}''
\cite{stachel}..., [but perhaps more importantly, that] ``{\small
[Quantum theory]  {\em does not seem to be in accordance with a
continuum theory, and must lead to an attempt to find a purely
algebraic theory for the description of reality. But nobody knows
how to obtain the basis of such a theory}'' \cite{einst3}}.
\end{quotation}

\noindent Indeed, we are tempted to say that when Einstein was
talking about ``{\em ...concepts which have proven useful for
ordering things easily assume so great an authority over us, that
we forget their terrestrial origin and accept them as unalterable
facts. They then become labelled as `conceptual necessities', `a
priori situations', etc.}'' in the quotation we saw in the
introduction, he was `subconsciously' referring to the {\it a
priori} concept (and use by CDG-means) of the spacetime continuum
in GR. Moreover, again to emulate Einstein's concluding words in
that quotation, we believe that

\begin{quotation}
\noindent{\em the road of progress in QG has been blocked for a
long period by our erroneous insistence on the `physicality' of
the background geometrical spacetime continuum}.
\end{quotation}

\noindent Parenthetically, and on more general grounds, let it be
stressed here that Einstein, during his later years, went as far
as to insist that (and we quote him indirectly via Peter Bergmann
from \cite{berg0}):\footnote{This quotation can also be found in
Anastasios Mallios' contribution to this volume.}

\begin{quotation}
\noindent ``{\small ...geometrization of physics is {\em not} a
foremost or even a meaningful objective...}''
\end{quotation}

\noindent Thus, we see that Einstein towards the end of his life
tended to leave behind `geometry' and take on `algebra' {\it
vis-\`a-vis} the quantum domain.

Lately, Einstein's words for a purely algebraic description of
physical phenomena in the quantum deep in the penultimate
quotation above, have found fertile ground as there have been
tendencies towards a purely algebraic theoresis of QG. Ten years
ago, Louis Crane asked characteristically in the very title of a
paper of his \cite{crane0}:

\vskip 0.1in

\centerline{\em ``Clock and category: is quantum gravity
algebraic?''}

\vskip 0.1in

\noindent The purely algebraico-categorical ADG-gravity appears to
answer to it affirmatively, and what's more, in a {\em background
spacetime manifoldless differential geometric setting}, in spite
of Isham's doubts and reservations above. May ADG provide the
theoretical framework that Einstein was (and some of us still are
nowadays!) looking for in our journey towards QG. However, even if
that does not turn out to be the case in the end, at least we will
have in our hands an entirely algebraic (re)formulation of
differential geometry---a novel framework pregnant with new
concepts, new principles, new techniques, and new theoretical
terms. Following Wallace Stevens' \cite{stevens} dictum, that:

\vskip 0.1in

\centerline{\small ``...Progress in any aspect is a movement
through changes in terminology...''}

\vskip 0.1in

\noindent we believe it is worth trying to move towards our QG
destination through the ADG-path.

Let us now pick the argument from where we left it earlier, when
the problem of gravitational singularities was discussed under the
prism of the background spacetime manifoldless ADG-gravity, and
comment on the closely related problem of the unphysical
infinities associated with those singularities, as well as the
non-renormalizable infinities appearing in QG when treated as
another, manifold based, QFT.

\paragraph{Whence the unphysical infinities?} There are infinities associated with gravitational
singularities, there is no doubt about that. For instance, the
curvature of the spherically symmetric Schwarzschild gravitational
field of a point-particle of mass $m$ diverges as
$\frac{m^{2}}{r^{6}}$ as one approaches it ($r\mapto 0$);
moreover, there is no analytic extension of the Schwarzschild
spacetime manifold so as to include the singular {\it locus} $m$
with the other regular points of the manifold \cite{clarke4}. In
contradistinction to the exterior Schwarzschild singularity at
$r=2m$ (:horizon) which has been branded a virtual, coordinate
singularity, the interior $r=0$ one is thought of as a true
singularity, with physical significance. Nevertheless, it is
altogether hard to believe that there are actually physically
meaningful infinities in Nature.

As noted earlier, many researchers hoped (and still do!) that QG
will remove singularities in the same way that QED removed the
Maxwellian infinities. Thus, perturbative QG, by emulating the
other quantum gauge theories of matter, initially regarded QG as
another QFT (on a flat Minkowski background!) and evoked the
(arguably {\it ad hoc}) process of renormalization to remove
gravitational infinities. It soon failed miserably, because of the
dimensionful gravitational coupling constant. Theoretical
physicists are people of resourcefulness, strong resolve and stout
heart, thus they evoked (or `better', they introduced by hand!)
extra dimensions, extra fields to occupy them and extra symmetries
between those extra fields ({\it eg}, supergravity and
supersymmetric string theories) in order to `smear' the offensive
{\it loci}, much like one blows up singularities in algebraic
geometry. The singular interaction point-vertices of the Feynman
diagrams of the meeting propagation lines of the point-particles
of QFT were smeared and smoothened out by world-tubes of
propagating closed strings, being `welded' smoothly into one
another at the interaction sites. However, infinities, although
tamed a bit, are still seen to persist galore (never mind the
grave expense of theoretical economy that accompanies the
introduction of more and more in principle unobservable fields and
their particle-quanta).\footnote{Of course, this blatant violation
of Occam's razor is not necessarily bad by itself, as at least it
keeps the experimentalists busy (and quiet!) designing experiments
to look for the `predicted' extra particles ({\it eg}, the
superpartners of the known particles), whose existence appears to
be mandated by theory.}

At the same time, people from the non-perturbative QG camp soon
realized that non-renormalizability is not a problem in itself if
one takes into consideration that QG, as opposed to the other
quantum gauge forces of matter, has associated with it a
fundamental space-time scale---the Planck length-time---which as
noted earlier is an expression involving the fundamental constants
of the three theories that are supposed to be merged into QG: $G$
from (Newtonian) gravity, $c$ from relativity, and $\hbar$ from
quantum mechanics. The Planck scale can then be thought of as
prohibiting {\em in principle} the integration down to infinitely
small spacetime distances; or dually in the perturbation
series/integrals, up to infinite momenergies. Non-perturbative QG
fundamentally assumes that spacetime is inherently cut-off
(:`regularized') by the Planck scale, so that below it the
continuum picture should be replaced by something more discrete
and quantum.

All this is well known and good. The infinities have not only kept
us occupied for a while, but they have provided us with a wealth
of new ideas and techniques in our struggle and strife to remove
them ({\it eg}, anomalies, spontaneous symmetry breaking, phase
changes, catastrophes and other critical phenomena, as well as the
renormalization group technology that goes hand in hand with them,
{\it etc.}) \cite{jackiw}. However, their stubborn persistence
makes us still abide by our main thesis here: it is indeed the
background smooth spacetime continuum, accommodating uncountably
infinite degrees of freedom of the fields which are modelled after
smooth functions on it (or $\otimes_{\smooth_{M}}$-tensors
thereof), that is responsible for all those pestilential
infinities. We must therefore give up {\em in principle} the
spacetime continuum (:manifold) and the usual Analysis (Calculus
or CDG) based on it, because they appear to miscarry in the QG
deep. In this line of thought we can metaphorically paraphrase
Evariste Galois':

\vskip 0.1in

\centerline{``{\small\em Les calcules sont
impracticables}'',\footnote{``{\em Calculations are
impractical}''.}}

\vskip 0.1in

\noindent and add that {\em the Differential Calculus, when
effectuated via the background geometrical spacetime continuum, is
an obstacle rather than a boon to QG research}. In turn, this
reminds us of Richard Feynman calling the usual differential
geometry ``{\em fancy schmanzy}'', doubting the up-front
geometrical interpretation of GR, and opting instead for a
combinatory-algebraic (diagrammatic-relational) scheme along
QFTheoretic lines for its quantization \cite{feyn2}.\footnote{See
especially the forward, titled ``{\itshape Quantum Gravity}'',
written by Brian Hatfield, giving a brief account of Feynman's
approach to QG. Hatfield argues there that Feynman not only felt
that the (differential) geometrical interpretation of gravity
`gets in the way of its quantization', but also that it masks its
fundamental gauge character.} Of course, Feynman's unsuccessful
attempt at quantizing gravity by applying the
perturbative-diagrammatic technology of QED is well documented.

At the same time, from the ADG-gravitational perspective we cannot
accept the non-perturbative QG's thesis that there is a
fundamental spacetime scale in Nature either, simply because there
is no spacetime in Nature to begin with. From our viewpoint, in a
Leibnizian sense, {\em `spacetime' is the (dynamical) objects that
comprise `it'; that is, the (dynamical) fields}. Accepting the
existence of a fundamental scale in Nature, above which Einstein's
equations hold, but below which the latter break down and another
set of equations (:those of the QG we are supposed to be after)
are in force, is analogous to accepting singularities as {\em
physical} entities. They both violate the universality of Physical
Law, and undermine the unity and autonomy of the gravitational
field.

That our calculations are plagued by infinities is more likely
because the usual Differential Calculus that {\em we} employ is
inextricably tied to a geometrical base spacetime continuum that
{\em we} assume up-front in the theory. Our manifold based
Analysis invites infinities by allowing for infinitary processes
(of divergence) relative the base topological continuum. On the
other hand, {\em there is no infinity in algebra}, and our purely
algebraic ADG-gravity suffers from no such unphysical pathologies.
It would be weird, or indeed comical(!), to even try to fathom
what would the meaning of the notion of `singularity' be in a
purely `pointless' and `space(tile)less' algebraico-categorical
setting like ours. For example, an attempt at the following
analogy produces funny thoughts:

\begin{quotation}
\noindent Does a singularity bend (or break!) the categorical
arrows (:connection field morphisms) in ADG-gravity in a way
analogous to how a point-electron is geometrically envisaged to
distort the Faraday lines of force of the electromagnetic field in
its vicinity? Then, {\it mutatis mutandis} for the gravitational
field lines of force strongly focusing towards a point-mass, as in
the case of the interior Schwarzschild (black hole) singularity.
\end{quotation}

\paragraph{New theoretical-mathematical framework for QG.}
To connect this epilogue back to the prologue like the proverbial
tail-biting serpent, in QG research the glaring absence of
comprehensive experiments and thus of reliable and concrete
experimental data to support and constrain theory-making is, at
least from a mathematical viewpoint, quite liberating. The
tentative, transient and speculative nature of the field invites
virtually unrestrained conceptual imagination, mathematical
creativity and wild philosophical wandering.

Even that most austere and critical of all 20th century
theoretical physicists, Wolfgang Pauli, said about the prospect of
quantizing the gravitational field \cite{pauli}:

\begin{quotation}
\noindent ``{\small ...Every theoretical possibility is a
potential route to success...[however, in this field] only he who
risks has a chance to succeed...}''\footnote{See also Feynman's
quotation in the introductory paper to this volume.}
\end{quotation}

\noindent Abiding by John Wheeler's dictum that `{\em more is
different}', the plethora of (mathematical) approaches to QG are
more than welcome (even if we coined it the seemingly derogatory
`{\em zoo}' in the prologue!), under the proviso that every now
and then unifying efforts are made to patch together the mosaic of
approaches to QG into a single---or at least to a
regular---pattern tapestry. This can be achieved for example by
occasionally leaving the worm's eye-view---as it were, the
`local', nitty-gritty problems and technical calculations of each
individual approach---and by trying to attain a `global'
conceptual, bird's eye-view of the field; one that at least tries
to make `dictionary correspondences', in both conceptual and
technical jargon, between different approaches. For Nature is
economical, and so must be our theories of Her---if not in
(mathematical) technicalities, at least conceptually.

On the other hand, Paul Dirac, more than 70 years ago
\cite{dirac3}, implored us to apply all our existing mathematical
arsenal, and even to invent and create new mathematics in order to
tackle the outstanding theoretical physics problems of the last
century---QG arguably being {\em the} central one that stubbornly
resists (re)solution in our times:\footnote{Quote borrowed from
fairly recent paper by Ludwig Faddeev \cite{faddeev}.}

\begin{quotation}
\noindent ``{\small ...The steady progress of physics requires for
its theoretical foundation a mathematics that gets continually
more advanced. This is only natural and to be expected. What,
however, was not expected by the scientific workers of the last
century was the particular form that the line of advancement of
the mathematics would take, namely, it was expected that the
mathematics would get more complicated, but would rest on a
permanent basis of axioms and definitions, while actually the
modern physical developments have required a mathematics that
continually shifts its foundation and gets more abstract...It
seems likely that this process of increasing abstraction will
continue in the future and that advance in physics is to be
associated with a continual modification and generalization of the
axioms at the base of mathematics rather than with logical
development of any one mathematical scheme on a fixed foundation.

There are at present fundamental problems in theoretical physics
awaiting solution [...]\footnote{At this point Dirac mentions a
couple of outstanding mathematical physics problems of his times,
which are hereby omitted.} the solution of which problems will
presumably require a more drastic revision of our fundamental
concepts than any that have gone before. Quite likely these
changes will be so great that it will be beyond the power of human
intelligence to get the necessary new ideas by direct attempt to
formulate the experimental data in mathematical terms. The
theoretical worker in the future will therefore have to proceed in
a more indirect way. {\em The most powerful method of advance that
can be suggested at present is to employ all the resources of pure
mathematics in attempts to perfect and generalise the mathematical
formalism that forms the existing basis of theoretical physics,
and after each success in this direction, to try to interpret the
new mathematical features in terms of physical
entities}\footnote{Our emphasis.}...}''
\end{quotation}

\noindent At the same time, however, there is this nagging little
voice at the back of every theoretical physicist's mind cautioning
her about the {\em New Maths Version of Murphy's Law}, maintaining
that

\begin{quotation}
\noindent {\em whenever there is a 50-50 chance that a new
mathematical theory applies to physics successfully, 9 times out
of 10 it turns out to fail},\footnote{A watered down version of
what David Finkelstein has coined the `{\em mathetic fallacy}' in
theoretical physics (private communication).}
\end{quotation}

\noindent notwithstanding Eugene Wigner's `{\em unreasonable
effectiveness of mathematics}'. In turn, this further evokes
forebodings of scepticism and fear, reminding her of Pauli's
(in)famous remark that ``{\em this theory is so bad, it's not even
wrong}''.

Nevertheless, it is the main position of this author that such
reservations and phobias have to be put aside in the dawn of the
new millennium, for in the end they only present inertia to, and
create an attitude of pessimism (invariably resulting to
indolence) in the development of theoretical physics. We have to
be innovative, adventurous and unconventional, perhaps even
iconoclastic,\footnote{See opening paper in this issue.} not only
about our technical-mathematical machinery, but also about the
conceptual and philosophical underpinnings of our fundamental
theories of Nature---with QG in particular, since it is arguably
the deepest of them all. Gerard 't Hooft put it succinctly in
\cite{thooft}:

\begin{quotation}
\noindent ``{\small ...The problems of quantum gravity are much
more than purely technical ones. They touch upon very essential
philosophical issues...}''
\end{quotation}

\noindent Thus, we should not inappreciably pass-by this unique
opportunity that QG is offering us: to bring together Physics and
Philosophy, thus reinstate the luster of `{\em Naturphilosophie}'
that theoretical physics seems to have lost in the last century,
predominantly due to its focusing on technical (:mathematical)
formalism, atrophizing at the same time important
conceptual/interpretational issues.

Ultimately, we should not be afraid of making mistakes, or fear
that our theories will come short of describing Nature completely,
because anyway, on the one hand the maths is our own free
intellectual creation\footnote{Recall from the quotes given above
Einstein referring to the (mathematical) concept of the
(spacetime) continuum as a `{\em mode by which \underline{we}
think}', as well as his warning us in general not to forget the
`{\em terrestrial origin}' of various concepts, no matter how
useful they may have been in the past.} (thus, we can take
responsibility for their shortcomings and blemishes, and rectify
them when necessary), while on the other, Physis is almost {\it de
facto} wiser than us. This simply goes to show that theoretical
physics is a never ending quest, and thus that our theories are in
a constant process of revision, refinement and extension.

To close this epilogue the way we started it, as Faddeev maintains
in \cite{faddeev} motivated by Dirac's remarks above,
theoretical/mathematical physics cannot---in fact it {\em should
not}---rely anymore on experiment for its progress. It should
become more and more autonomous, more and more abstract, as well
as versatile and wide ranging. Once again, the tried and tested
age-old virtues of conceptual simplicity, mathematical economy and
beauty---virtues that are trademarks in the celebrated works of
such giants as Einstein and Dirac---can be called to guide us in
our theoretical physics (ad)ventures through our presumed
`subject': Physis.\footnote{Of course, if anything, {\em we} are
the subjects of Nature, not the other way round. Hence the
quotation marks.} And we can rest assured that these virtues shall
safeguard us from `mathematically arbitrary' theory-making.

\begin{quotation}
\noindent After all, it is well known that when the solar eclipse
results were due back from Arthur Eddington's 1919 Cape Town
expedition, in Berlin Max Planck could not go to sleep in
anticipation and excitement about whether GR would be
experimentally (:observationally) vindicated; or on the contrary,
whether it would fail to deliver in the end. {\em Einstein on the
other hand reportedly went to bed by eight o'clock}...
\end{quotation}

\section*{Acknowledgments}

This author is indebted to the following people for numerous
exchanges, feedback and critique, as well as for their moral
encouragement and material support over the past half-decade, to
his research program of applying ADG to classical and quantum
gravity: Chris Isham, Jim Lambek, Tasos Mallios, Chris Mulvey,
Steve Selesnick, Rafael Sorkin, John Stachel, Petros Wallden and
Roman Zapatrin. He would also like to acknowledge financial
support from the European Commission in the form of a European
Reintegration Grant (ERG-CT-505432) held at the University of
Athens (Greece).

\end{document}